\documentclass[conference]{IEEEtran}
\IEEEoverridecommandlockouts

\usepackage{cite}
\usepackage{amsmath,amssymb,amsfonts}
\usepackage{graphicx}
\usepackage{textcomp}
\def\BibTeX{{\rm B\kern-.05em{\sc i\kern-.025em b}\kern-.08em
    T\kern-.1667em\lower.7ex\hbox{E}\kern-.125emX}}

\usepackage{tikz}
\usepackage{filecontents}
\usepackage{multicol}
\usepackage{multirow}
\usepackage{lipsum}
\usepackage{newfloat}
\usepackage{listings}
\usepackage{tabularx}
\usepackage{makecell}
\usepackage{enumitem}
\usepackage{subcaption}
\usepackage{array}
\usepackage{caption}
\usepackage{float}
\usepackage{bm} 
\usepackage{url}

\usepackage{siunitx}
\usepackage{nicefrac} 
\usepackage{microtype} 
\usepackage{titlesec}
\usepackage{longtable}
\usepackage{hyperref}

\usepackage[most]{tcolorbox}
\usepackage{tabularx}
\usepackage{booktabs}
\usepackage{colortbl}
\usepackage{seqsplit}
\usepackage{pifont}

\usepackage[table]{xcolor}

\usepackage{algorithm} 
\usepackage{algpseudocode} 
  
\definecolor{TADgreen}{HTML}{2F7D5B}
\definecolor{TADgreenLight}{HTML}{EDF7F1}
\definecolor{TADgreenRow}{HTML}{E4F3EA}
\definecolor{SoftGray}{HTML}{F5F6F7}
\definecolor{MidGray}{HTML}{6B7280}
\definecolor{SoftOrange}{HTML}{FFF4E5}
\definecolor{OrangeText}{HTML}{A15C00}
\definecolor{SoftRed}{HTML}{FBEDEE}
\definecolor{RedText}{HTML}{A33A3A}
\definecolor{RuleGray}{HTML}{D1D5DB}
\definecolor{HeaderBlack}{HTML}{22252A}

\newcommand{\cmark}{\textcolor{TADgreen}{\ding{51}}}
\newcommand{\xmark}{\textcolor{RedText}{\ding{55}}}
\newcommand{\warnmark}{\textcolor{OrangeText}{\raisebox{0.1ex}{\scriptsize$\blacktriangle$}}}

\newtcolorbox{mybox}[2][]{
    enhanced,
    breakable=false,
    colback=gray!4!white,
    colframe=black!70,
    colbacktitle=black!75,
    coltitle=white,
    fonttitle=\bfseries,
    title=#2,
    boxrule=0.6pt,
    arc=2mm,
    left=3mm,
    right=3mm,
    top=2mm,
    bottom=2mm,
    before skip=8pt,
    after skip=8pt,
    attach boxed title to top left={
        xshift=3mm,
        yshift*=-\tcboxedtitleheight/2
    },
    boxed title style={
        colback=black!75,
        colframe=black!75,
        arc=1.5mm,
        boxrule=0pt,
        left=2mm,
        right=2mm
    },
    #1
}

\lstdefinestyle{logstyle}{
    basicstyle=\ttfamily\footnotesize,
    breaklines=true,
    columns=fullflexible,
    keepspaces=true,
    showstringspaces=false,
    frame=none,
    backgroundcolor=\color{gray!8},
    xleftmargin=2mm,
    xrightmargin=2mm,
    aboveskip=1mm,
    belowskip=1mm
}

\newtcolorbox{promptbox}{
  colback=gray!5!white, 
  colframe=gray!75!black,
  rounded corners,
  boxrule=0.3pt,
  coltitle=black,
}

\begin{document}

\title{Topological Attribution Distance (TAD): \\ Revealing Segment-Level RAG Influence on LLM Output Geometry for Incident Log Analysis}

\author{\IEEEauthorblockN{1\textsuperscript{st} Reza Fayyazi}
\IEEEauthorblockA{\textit{Dept. of Computer Engineering} \\
\textit{Rochester Institute of Technology}\\
Rochester, NY, USA \\
rf1679@rit.edu}
\and
\IEEEauthorblockN{2\textsuperscript{nd} Michael Zuzak}
\IEEEauthorblockA{\textit{Dept. of Computer Engineering} \\
\textit{Rochester Institute of Technology}\\
Rochester, NY, USA \\
mjzeec@rit.edu}
\and
\IEEEauthorblockN{3\textsuperscript{rd} Shanchieh Jay Yang}
\IEEEauthorblockA{\textit{Dept. of Engineering} \\
\textit{Gonzaga University}\\
Spokane, WA, USA \\
yangj@gonzaga.edu}
}

\maketitle

\begin{abstract}
 Large Language Models (LLMs) are increasingly being deployed in cybersecurity operations to assist cybersecurity analysts with rapid decision-making against emerging threats. However, there is a main criteria that must be met when using LLMs in cybersecurity, that is, \underline{trust} in the generated outputs. As Agentic AI is integrated into operational systems, a robust evidence attribution and provenance tracking technique is essential to \underline{trace} the origins of model generations. When autonomous agents make a decision (right or wrong), the ability to trace back through the decision chain is critical, as without it, teams cannot identify which segment of the data caused the model generation. Existing methods often struggle to distinguish among complex and highly similar evidence sources, such as cyber incident logs. This reveals a key gap: current approaches do not adequately capture the holistic geometric relationship between the retrieved evidence and the generated response for reliable evidence verification. To bridge this gap, we propose \textbf{Topological Attribution Distance (TAD)}, inspired by Topology, to characterize and capture the global geometric shape of an output and its changes against its retrieved logs. In other words, if the embeddings of a specific source log drastically changes the geometry of the model's response in the embedding space, this suggests that such log is a critical source for the model's generated response. Therefore, TAD is powered by segment-level ablation attribution to investigate incident logs of an actual cyberattack. We demonstrate how TAD finds the most attributed logs on LLM outputs in an adaptive manner. This can provide an explainable and trustworthy tracing based on each LLM's hidden state to understand how geometrically different retrieved logs influence the model generation, and provide evidence verification in cybersecurity and Agentic-AI workflows.
\end{abstract}


\section{Introduction}

Retrieval-Augmented Generation (RAG) has become a widely used approach in Agentic-AI workflows by grounding Large Language Models (LLMs) with relevant external knowledge. In cybersecurity, RAG allows LLMs to access up-to-date information, such as vulnerability reports and incident logs, to help support more informed security analysis and decision-making. However, retrieval alone does not guarantee correct interpretation, particularly when models are exposed to large pools of highly structured and semantically similar information. As noted by Zhang et al. \cite{zhang2025hallucination}, RAG systems remain susceptible to ungrounded outputs when queries are ambiguous, retrieved context is excessive, or source quality is poor. This raises a critical concern for AI adoption in cybersecurity applications, and that is: \emph{trust} in the generated outputs. As Agentic AI becomes integrated into operational systems, robust evidence attribution and provenance tracking are essential for tracing model decisions back to their originating sources. Without such traceability, decisions made by AI agents (whether right or wrong) are difficult to diagnose, verify, and audit. Therefore, there remains the challenge to reliably attributing model decisions to their originating sources for evidence verification. Understanding why a model generated a particular response and which retrieved context influenced the models' decision is critical for providing trust, transparency, and auditability in the age of Agentic AI. 

To tackle these challenges, there have been several lines of work that investigated providing explainability and attribution for LLMs' generated responses \cite{selfrag, es2024ragas, saad2024ares, fayyazi2025llm, hu5199254lrp4rag, paes2025multi}, yet each carries notable limitations. One prominent approach employs LLM-based judges to assess provenance \cite{selfrag, es2024ragas, saad2024ares}. Because these methods rely on the same class or another set of models being evaluated, their accuracy is fundamentally constrained by the biases, hallucination tendencies, and self-assessment patterns inherent to LLMs themselves. A second body of work focuses on token-level attribution by producing importance scores for individual input tokens to identify which tokens influence a given generation the most \cite{fayyazi2025llm, hu5199254lrp4rag, paes2025multi}, or similarity measures (e.g., cosine similarity and Rouge-L). These methods are either computationally expensive or ineffective on highly structured and similar data, such as incident logs. Therefore, these methods do not capture \emph{holistic segment-level grounding}, which is the degree to which a \emph{model's response as a whole} is driven by retrieved sources. In other words, a security log's influence on model behavior is better understood holistically, as a complete unit of evidence, rather than as a collection of individually scored tokens. For example, in structurally similar logs, the key distinction may arise from the connectivity between a few tokens, such as \texttt{psexec.exe} appearing at an unusual timestamp. Token-level attribution can dilute such signals across many shared, non-discriminative tokens. This highlights a key gap in current attribution methods: the lack of a mechanism that aggregates token-level signals into coherent segment-level attribution that identifies when a small but critical feature makes an entire retrieved segment influential.

To account for these challenges, we turn into Topological Data Analysis (TDA), a mathematical framework for studying the \emph{shape} of data \cite{chazal2021introduction}. TDA draws on concepts from Topology, where algebraic invariants, such as homology and homotopy groups, are assigned to topological spaces to characterize their structure \cite{munkres2025elements}. Homotopy and homology are two lenses on the same underlying question: ``When are two topological spaces the same''? Homotopy is a continuous deformation between two maps ($f, g: X \to Y$), formalized as a continuous function ($H: X \times [0, 1] \to Y$) satisfying ($H(x, 0) = f(x)$) and ($H(x, 1) = g(x)$). Intuitively, homotopy captures the idea that two spaces are equivalent if one can be continuously stretched or compressed into the other. While homotopy is the more intuitive notion, it is generally intractable to compute directly \cite{anick_homotopy}. Homology addresses this by extracting computable algebraic invariants (e.g., H0 for connected components, H1 for finding holes, etc.) that are preserved under homotopy equivalence. In other words, homology acts as a computationally tractable proxy for homotopy, meaning that if two spaces have different homology, they cannot be homotopy-equivalent. 

Existing research has shown that feed-forward networks progressively transforms the topology of the data as they flow through the layers, which means that Topology has the capability to map the non-linear space \cite{paluzo2025latent}. Prior work has also shown that deeper layer does not always mean better representation, demonstrating the importance of effect of all the layers into the final conclusion \cite{harun2024variables}. Therefore, in the context of LLMs, the geometric and topological properties of the embedding spaces over the layers (i.e., non-linear hidden-state spaces) can produce meaningful explanations about model decisions, attribution, and uncertainty \cite{gardinazzi2025persistent, fay2026shape}. Interestingly, the auto-regressive nature of LLMs (i.e., prior context shaping the response) can provide a powerful lens for reasoning about model generations. Instead of comparing isolated token representations from the response, we can characterize the \emph{token connectivity of a response as a whole} through the embedding space and trace the influence of each prior context (each segment) individually. In other words, if the inclusion and omission of a retrieved context embeddings can strongly deform the geometry of the model's response, this can suggest as evidence that the context served as a grounding source. This topological perspective directly addresses the segment-level attribution gap identified above. 

Building on this intuition, we propose \textbf{Topological Attribution Distance (TAD)}, a segment-level attribution metric to identify the segments (e.g., logs) that cause the most geometric change in an LLM's generation. TAD operates by tracking the topological signatures of the generated response's embedding geometry over the transformer layers, both with and without each retrieved segment present. More specifically, we compute the generated response's persistence diagram from their embedding geometries across each transformer layer, with and without each segment in the context, and quantify the changes on the response embeddings with Wasserstein distance. By measuring the topological cost of significant changes on a generated response's geometry, TAD produces an attribution score at the segment-level. This score captures the degree to which each retrieved segment geometrically influenced the model's output.

We evaluate TAD on a real-world cyberattack logs, and on three different scenarios, namely \emph{Direct}, \emph{Regular}, and \emph{Indirect}. These scenarios point to the level of difficulty of a response with respect to the ground-truth attack logs. More specifically, the \emph{Direct} case contains many tokens from the log in the response, the \emph{Regular} case contains some specific keywords, and the \emph{Indirect} case contains little-to-none token overlaps between the logs and the response. Moreover, we compare TAD against multiple baselines, namely LLM-as-a-judge, similarity measures, and token-level attribution techniques. We demonstrate the effectiveness of the proposed TAD metric across four different LLMs (with different sizes and architectures). Our results demonstrate that TAD significantly outperforms existing baselines with over 97\% accuracy on average in tracing the top-attributed logs, and achieving over 39\% gap in Precision compared to the best-performing baseline in all three scenarios.  In Figure \ref{fig:tad-motivating-case-study}, we provide a case study to demonstrate how different attribution techniques fail in tracing security logs to the response, and how TAD works well in these cases. Together, these results position TAD as an auditable, provenance-aware measure capable of tracing model decisions back to their retrieved sources, and therefore, building trust in cybersecurity operations and critical decision-making.

The key contributions of this work are as follows:
\begin{itemize} 
\item We formalize the problem of segment-level attribution in RAG systems as a topological comparison problem, drawing an explicit connection between homotopy-theoretic reasoning and embedding-space geometry. 
\item We introduce \textbf{Topological Attribution Distance (TAD)}, a novel attribution metric that captures segment-level influence of changes in LLM response geometry over the hidden states, and compute their Wasserstein distance to find the top-attributed segments. 
\item We demonstrate that TAD is well-suited for incident log analysis, as logs share high token overlap within each other, and show how TAD is capable of identifying the logs that influence the LLM response's geometry the most.
\item We demonstrate that the proposed TAD metric outperforms other baselines with an average of 97\% accuracy on four different LLMs, and with over 39\% gap in Precision for the three different difficulty-level cases.
\end{itemize}

\begin{figure*}[t]
\centering

\begin{tcolorbox}[
    enhanced,
    width=\textwidth,
    colback=white,
    colframe=RuleGray,
    boxrule=0.4pt,
    arc=1.2mm,
    left=2mm,
    right=2mm,
    top=1.1mm,
    bottom=1.1mm,
    title=\textbf{Case Study: From an LLM Generation to Tracing Back its Supporting Evidence},
    fonttitle=\normalsize\bfseries,
    colbacktitle=HeaderBlack,
    coltitle=white,
    titlerule=0pt,
    toptitle=1mm,
    bottomtitle=1mm
]

\footnotesize
\setlength{\tabcolsep}{3.2pt}
\renewcommand{\arraystretch}{1.05}

\noindent\textbf{\underline{1. Analyst query.}}

\begin{tcolorbox}[
    enhanced,
    width=\linewidth,
    colback=SoftGray,
    colframe=RuleGray,
    boxrule=0.3pt,
    arc=0.8mm,
    left=1.4mm,
    right=1.4mm,
    top=0.55mm,
    bottom=0.55mm
]
\footnotesize \centering
Examine logs around 00:01--00:05 on 2025-01-01.
Exactly one of the following five logs is evidence of malicious activity.
Identify its \texttt{log\_id} and explain why it is malicious in one sentence.
\end{tcolorbox}

\vspace{0.7mm}

\noindent\textbf{2. Candidate incident logs}

\vspace{0.3mm}

\begin{tabularx}{\linewidth}{
    >{\centering\arraybackslash}p{0.045\linewidth}
    >{\centering\arraybackslash}p{0.20\linewidth}
    >{\raggedright\arraybackslash}X
    >{\centering\arraybackslash}p{0.11\linewidth}
}
\toprule
\rowcolor{SoftGray}
\textbf{Log} & \textbf{Log-ID} & \textbf{Event} & \textbf{State} \\
\midrule

1 &
2e9a4...5a094 &
\texttt{jdoe} starts \texttt{teams.exe}. &
\textcolor{MidGray}{Benign} \\

\rowcolor{SoftGray}
2 &
8b1d6...fa5bd &
\texttt{SYSTEM} modifies
\texttt{HKLM\textbackslash...\textbackslash WindowsUpdate}. &
\textcolor{MidGray}{Benign} \\

3 &
d0c4a...73f06b &
\texttt{jdoe} logs on via \texttt{explorer.exe}. &
\textcolor{MidGray}{Benign} \\

\rowcolor{SoftGray}
4 &
3a7e1...bd5a02 &
\texttt{jdoe} starts \texttt{code.exe}. &
\textcolor{MidGray}{Benign} \\

\rowcolor{TADgreenRow}
\textbf{\textcolor{TADgreen}{5}} &
\textbf{7c3f0...b0a48} &
\textbf{\texttt{jdoe} writes
\texttt{HKCU\textbackslash...\textbackslash Run}
$\rightarrow$
\texttt{\%TEMP\%\textbackslash svc.exe}.} &
\textbf{\textcolor{TADgreen}{Malicious}} \\

\bottomrule
\end{tabularx}

\vspace{0.7mm}

\vspace{2mm}

\noindent\textbf{\underline{3. LLM Response (Gemma3-4b).}}

\begin{tcolorbox}[
    enhanced,
    width=\linewidth,
    colback=SoftGray,
    colframe=RuleGray,
    boxrule=0.3pt,
    arc=0.8mm,
    left=1.4mm,
    right=1.4mm,
    top=0.55mm,
    bottom=0.55mm
]
\scriptsize
``Log ID \texttt{7c3f0a9e2b514d68} is the indicator of malicious activity, as the system initiated the launch of an executable from a temporary location, suggesting an attempt to establish covert operation.''

\end{tcolorbox}

\begin{center}
\textbf{\underline{Observed problem.}
The model identifies the correct event but partially hallucinates its exact log ID, requiring evidence localization.}
\end{center}

\vspace{0.0mm}

\begin{center}
\footnotesize
\textbf{\textit{$\Downarrow$ Can the Supporting Evidence be Verified?}}
\end{center}

\vspace{0.25mm}

\noindent\textbf{4. Evidence localization}

\vspace{0.5mm}

\begin{tabularx}{\linewidth}{
    >{\raggedright\arraybackslash}p{0.125\linewidth}
    >{\raggedright\arraybackslash}p{0.09\linewidth}
    >{\centering\arraybackslash}p{0.115\linewidth}
    >{\raggedright\arraybackslash}X
}
\toprule
\rowcolor{SoftGray}
\textbf{Family} &
\textbf{Method} &
\textbf{Result} &
\textbf{Operational consequence} \\
\midrule

\multirow{2}{*}{Similarity}
&
Cosine
&
\xmark\ Log 4
&
Retrieves a semantically similar but benign process event. \\

\cmidrule(lr){2-4}

&
ROUGE-L
&
\xmark\ Log 2
&
Lexical overlap favors a benign registry event. \\

\midrule

\multirow{2}{*}{Self-eval.}

&
Citation
&
\xmark\ Logs 1,3,4,5
&
Includes the true log but also incorrectly cites multiple benign events. \\

\cmidrule(lr){2-4}

&
LLM judge
&
\warnmark\ Logs 4,5
&
Includes the true log but leaves a benign candidate for manual review. \\

\midrule

Token-level attribution
&
LEA \cite{fayyazi2025llm}
&
\xmark\ Log 3
&
Ranks a benign logon event first as logs share so many overlapping tokens. \\

\midrule

\rowcolor{TADgreenRow}
\textbf{Segment-level attribution}
&
\textbf{TAD (ours)}
&
\cmark\ \textbf{Log 5}
&
\textbf{Uniquely identifies the log supporting the response; verifying evidence localization.} \\

\bottomrule
\end{tabularx}

\end{tcolorbox}

\caption{
Evidence-localization case study. The Gemma model identifies the malicious registry-persistence event but partially hallucinates its exact log identifier, making the generated answer insufficient for direct verification. Existing methods retrieve benign logs, return multiple candidates, or cite almost the full context. TAD uniquely localizes Log-5 by measuring the effect on the response geometry, allowing an analyst to trace the response to its supporting evidence.
}
\label{fig:tad-motivating-case-study}

\end{figure*}

\section{Related Works}

\subsection{Challenges of Existing Attribution Metrics for LLMs}
\label{sec: current_att_limitation}

The emergence of RAG systems has led to a line of research on source attribution, which is verifying whether the generated text is supported by identified source documents. For LLM-based judges, there have been different works proposed to evaluate RAG systems. Self-RAG \cite{selfrag}, RAGAS \cite{es2024ragas} and ARES \cite{saad2024ares} employ LLM-based judges to self-assess RAG systems, either through the generation model itself, or through another set of LLMs. However, these methods are constrained by the evaluation models' self-interpretation and reasoning biases, and they do not fully decouple attribution assessment from parametric model knowledge (i.e., hidden states).

Token-attribution methods aim to quantify the contribution of individual input features, typically token embeddings, on a model’s output. Numerous works have addressed the problem of attributing model predictions to individual input features \cite{li2023survey}. In LLMs, LIME \cite{ribeiro2016should} generates variations of the input text (e.g., by masking or removing tokens) and observes changes in output probabilities \cite{paes2025multi}, and SHAP \cite{lundberg2017unified} scores are derived by masking or perturbing subsets of input tokens and observing output variation \cite{paes2025multi}. However, these methods are computationally expensive, with SHAP being exponential, and cannot scale with a large pool of retrieved context (e.g., security logs). LRP \cite{bach2015pixel} is a backpropagation-based method that redistributes the model's prediction score backward through the network layers, assigning relevance scores to input features, and there have been works used LRP for attribution \cite{achtibat2024attnlrp, hu5199254lrp4rag}. Furthermore, in our previous work \cite{fayyazi2025llm}, we proposed the LLM Embedding-based Attribution (LEA) metric, a computationally efficient token-level attribution metric based on linear dependency analysis of input embeddings. Overall, despite significant progress, existing feature attribution techniques are limited to the \textbf{token-level} and can struggle when retrieved information contains highly overlapping tokens (e.g., logs). These methods  While token-level attribution is useful for identifying which tokens most influence a given prediction, it does not capture holistic segment-level grounding, which is the degree to which the \emph{model's response as a whole} is driven by the retrieved context. This points to a broader open challenge in the field: bridging the gap between token-level attribution into segment-level attribution tracing in RAG systems.

\subsection{Topological Data Analysis (TDA) in LLMs}

The application of Topological Data Analysis to language models spans several interrelated directions \cite{sekuloski2026exploring}. Draganov and Skiena \cite{draganov2024shape} demonstrate that persistent homology applied to word embedding clouds encodes meaningful linguistic structure. Rottach et al. \cite{rottach2025topology} introduced Unified Topological Signatures, which is a framework that aggregates topological and geometric descriptors to characterize embedding spaces. The authors revealed that models within the same family show similar topological properties, which suggests that architecture and training data strongly shape the embedding space geometry.

Within LLM internal representations, there are some works that apply topology directly to hidden-state activations. Fitz et al. \cite{fitz2024hidden} compare the layer-wise topological complexity of Transformer and LSTM hidden-states using persistent homology and perforation, and they found that LSTMs exhibit richer topological structure specific to natural language. Gardinazzi et al. \cite{gardinazzi2025persistent} employ zigzag persistence to track how topological features evolve across layers, introducing a Persistence Similarity metric that identifies redundant layers and allows principled pruning with minimal performance loss. Fay et al. \cite{fay2026shape} propose persistent homology as an architecture-agnostic adversarial detection framework, identifying a consistent ``topological compression'' signature under prompt injection and backdoor attacks, where adversarial inputs simplify latent-space structure, and demonstrating clean separation of normal and adversarial activations via barcode summary statistics.

Despite their promise, existing topological approaches to LLM analysis have notable limitations. Fitz et al. \cite{fitz2024hidden} demonstrated transformer models does not show topological complexity as going deeper in the layers (compared with LSTM). However, their perforation metric excludes $H0$ homology to capture connected components. We will show in this paper, that H0 plays a key role in finding shape differences in the embedding space. Gardinazzi et al. \cite{gardinazzi2025persistent} showed the importance of each layer in different LLMs. While certain layers contribute more significantly than others to model outputs, pruning even the least critical layers resulted in substantial degradation in benchmark performance. This indicates the importance of representation across all layers, and therefore, motivating us to incorporate this insight into our proposed methodology. Furthermore, Fay et al. \cite{fay2026shape} identified topological compression as an adversarial signature across six LLMs. However, the framework focuses exclusively on a last-token hidden state and relies on batch-level statistics, which requires them to depend on batches of data on similar categories (clean or poisoned) to obtain barcode summaries. This dependence on homogeneous batches to define relational geometry prevents the attribution of a single generated response in isolation, and lacks real-time applicability. Finally, none of these works account for segment-level attribution and multi-source influence on response generation.

\section{Background on Topological Data Analysis}
\label{sec: TDA_background}
In this section, we explain the mathematical foundations of \emph{Topological Data Analysis (TDA)}. We discuss the construction of simplicial complexes from point cloud data, persistent homology to track topological features across scales, and the Wasserstein distance for comparing persistence diagrams.

\subsection{Point Clouds and Metric Spaces}

The input to TDA is typically a point cloud, i.e., a finite set represented as:
\begin{equation}
   V = \{v_1, v_2, \ldots, v_n\} \subseteq M,
\end{equation}
where each point $v_i$ is embedded in a feature space $M$ equipped with a metric $d : M \times M \to \mathbb{R}$, so that $(M, d)$ forms a \textbf{metric space}.  The metric $d(v_i, v_j)$ quantifies the dissimilarity between any two points.  A common choice when $M = \mathbb{R}^n$ is the $L_\infty$ distance,
\begin{equation}
    d(v, w) = \|v - w\|_\infty = \max_{k}\,|v_k - w_k|.
\end{equation}
The central insight of TDA is that a discrete set of points carries latent geometric and topological information that can be systematically extracted from their pairwise distances  \cite{carlsson2009topology}. Thus, doing this requires lifting the point cloud to a higher-level combinatorial structure called a simplicial complex.

\subsection{Simplicial Complexes}

A simplicial complex provides a combinatorial representation of a point cloud using basic building blocks called simplices \cite{edelsbrunner2002topological}.  A $k$-simplex is defined as the convex hull of $k+1$ affinely independent points.  In other words, $k = 0$ gives vertices, $k = 1$ gives edges, $k = 2$ gives filled triangles, etc.
One standard construction for building a simplicial complex from a metric point cloud is the \textbf{Vietoris--Rips complex} \cite{hausmann1995vietoris}.  Given a scale parameter $\varepsilon > 0$, the Vietoris--Rips complex over a point set $V$ is defined as:
\begin{equation}
    \mathcal{R}_\varepsilon
    = \bigl\{ \sigma \subseteq V \;\big|\;
      d(v_i, v_j) \leq \varepsilon \;\;\forall\, v_i, v_j \in \sigma \bigr\}.
\end{equation}
That is, a $k$-simplex is included whenever \emph{all} pairwise distances among its $k+1$ vertices are at most $\varepsilon$.  The choice of $\varepsilon$ decides which connections are formed: too small $a$ value causes an almost empty complex capturing only trivial local structure, while too large $a$ value collapses all points into a single fully connected component, which motivates studying the complex across all scales simultaneously.

\subsection{Filtrations and Multi-Scale Analysis}

There is no value of $\varepsilon$ that can be globally optimal.  TDA resolves this ambiguity by considering \emph{all scales simultaneously} through a \textbf{filtration} \cite{edelsbrunner2002topological}, which is a nested sequence of simplicial complexes:
\begin{equation}
      \emptyset = K_0 \subseteq K_1 \subseteq K_2 \subseteq \cdots
      \subseteq K_m = K,
\end{equation}
These are indexed by an increasing sequence of parameter values $\varepsilon_0 < \varepsilon_1 < \cdots < \varepsilon_m$. As $\varepsilon$ grows, new simplices are progressively added. Initially, points appear as isolated vertices; edges then form between nearby points; eventually, higher-dimensional simplices emerge, which represents clustered regions of the space. Given a scalar function $f : \mathcal{V} \to \mathbb{R}$ defined on the point cloud $\mathcal{V} \subset \mathbb{R}^n$, the associated \textbf{sublevel-set filtration}  \cite{cohen2005stability} is the nested family of subsets
\begin{equation}
    \mathcal{V}_t = f^{-1}\!\bigl((-\infty,\, t]\bigr)
    = \bigl\{\, v \in \mathcal{V} \;\big|\; f(v) \leq t \,\bigr\},
\end{equation}
which is obtained by progressively admitting points in order of increasing the $f$-value as the threshold $t$ grows.  At each step, a simplicial complex (e.g., a Vietoris--Rips complex) is built on $\mathcal{V}_t$, capturing how the topology of the space evolves as points are included according to their functional values.

At each stage of the filtration, the shape of the complex is quantified using \textbf{homology} \cite{munkres2025elements}.  For each non-negative integer $k$, the \textbf{$k$-th homology group} $H_k(K)$, typically computed over a field $\mathbb{F}$ (most often $\mathbb{F}_2 = \{0,1\}$), captures independent topological features of dimension $k$. More specifically, $H_0(K)$ counts connected components, $H_1(K)$ captures independent loops or cycles, $H_2(K)$ detects enclosed voids, and higher $H_k(K)$ encode their higher-dimensional analogues.

\subsection{Persistent Homology: Birth, Death, and Persistence}

Persistent Homology tracks how homological features evolve across a filtration \cite{carlsson2009topology, edelsbrunner2002topological, zomorodian2004computing, munch2013applications}.  As the filtration parameter $\varepsilon$ increases, topological features \emph{are born} (first appear in $H_k(K_\varepsilon)$) and eventually \emph{die} (merge with older components or are filled in by higher-dimensional simplices). The
\textbf{persistence} of the feature is:

\begin{equation}
      \operatorname{pers}(\alpha) = \varepsilon_{death} - \varepsilon_{birth}.
\end{equation}

Features with high persistence are considered topologically significant, while features with low persistence are typically attributed to noise or sampling variability \cite{cohen2005stability}. Note that at least one connected component in $H_0$ never dies, since some component persists for the entire filtration. The complete lifecycle of topological features can be summarized in two
equivalent representations:

\noindent \textbf{Persistence Barcode} \cite{ghrist2008barcodes, carlsson2004persistence}: A persistence barcode is a multiset of intervals $[b_i, d_i)$, drawn as horizontal bars, where each bar corresponds to one topological feature and its length equals the feature's persistence.  Long bars indicate stable features and short bars (near zero) indicate a noisy structure.

\noindent \textbf{Persistence Diagram} \cite{edelsbrunner2002topological, cohen2005stability, chazal2009proximity}: A persistence diagram is a multiset of points $(b_i, d_i) \in
\overline{\mathbb{R}}^2$ with $d_i \geq b_i$, plotted in the half-plane above the diagonal $\Delta = \{(x,x) : x \in \mathbb{R}\}$.  The diagonal itself is included with infinite multiplicity, representing features that die immediately upon birth.  Points far from $\Delta$ correspond to significant features, while points near $\Delta$ indicate noise.  The persistence diagram arising from a filtration function $f$ in dimension $k$ is denoted as \underline{$\operatorname{Dgm}_k(f)$}. Therefore, having established a means of representing topological features as persistence diagrams, a natural question arises: ``how does one quantify the similarity (or distance) between two such diagrams in a geometrically meaningful way?'' We will discuss this next.

\subsection{Wasserstein Distance}

To compare two persistence diagrams $P$ and $Q$, which may have different  cardinalities, one seeks a \textbf{partial matching} $\gamma$ that minimizes a total cost \cite{cohen2005stability, chazal2009proximity, mileyko2011probability}. To accommodate unmatched points, let $\widehat{P} = P \cup \Delta$ and  $\widehat{Q} = Q \cup \Delta$ denote the diagrams augmented with the diagonal $\Delta = \{(x, x) : x \in \mathbb{R}\}$, treated as trivial features with zero persistence. A \textit{matching} is then a bijection $\pi : \widehat{P} \to \widehat{Q}$, where any point paired with  a diagonal point is considered \textit{unmatched}. The cost of transporting  a point $u = (b_u, d_u)$ to a point $v = (b_v, d_v)$ under the  $L^\infty$ norm is:
\begin{equation}
    \|u - v\|_\infty = \max\!\bigl(|b_u - b_v|,\, |d_u - d_v|\bigr),
\end{equation}
while matching $u = (b, d)$ to its orthogonal projection onto $\Delta$ incurs the persistence-proportional cost
\begin{equation}
    \|u - \Delta\|_\infty = \tfrac{d - b}{2},
\end{equation}
reflecting the minimal $L^\infty$ displacement required to remove a topological feature of persistence $d - b$. Therefore, the \textbf{$(\infty, 1)$-Wasserstein distance} ($W_1^{(\infty)}$) between persistence diagrams $P$ and $Q$ is defined as:

\begin{equation}
\begin{aligned}
    \label{equ: wasserstein}
  W_1^{(\infty)}(P,\, Q)
  \;=\;
  \inf_{\gamma}
\Bigg[
    \sum_{(u,v)\,\in\,\gamma}
    \|u - v\|_\infty
    \;+\;
    \\
    \sum_{u\,\in\, P \setminus \gamma}
    d(u,\Delta) 
    \;+\;
    \sum_{v\,\in\, Q \setminus \gamma}
    d(v,\Delta)
\Bigg]
  \end{aligned}
\end{equation}

where $\gamma$ ranges over all partial matchings between $P$ and $Q$, and  $d(u, \Delta) = (d_u - b_u)/2$ denotes the $\ell^\infty$-distance from  $u = (b_u, d_u)$ to its orthogonal projection onto $\Delta$. Intuitively, the first sum penalizes geometric discrepancy between matched features, while the second and third sums penalize unmatched features in $P$ and $Q$, respectively. 
Overall, Wasserstein distance provides a geometrically meaningful and computationally tractable measure of dissimilarity between persistence diagrams \cite{mileyko2011probability, cohen2010lipschitz, arulandu2025through}. This quantifies how much topological structure must be changed to transform one diagram into another.

\begin{figure*}[t]
\centering
\includegraphics[scale=0.395]{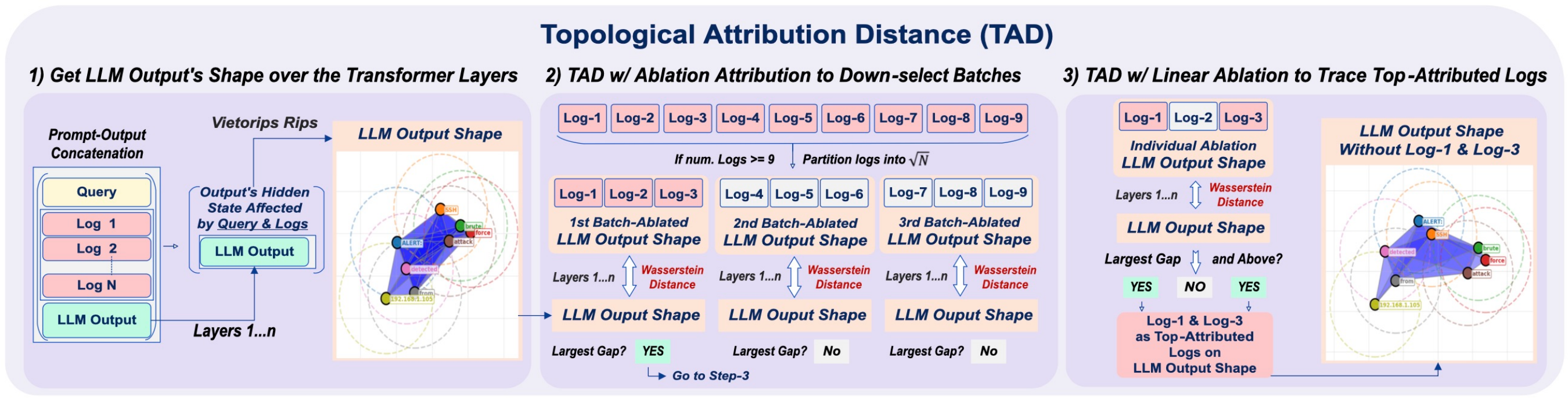}
\caption{TAD's procedure of tracing attribution with corresponding retrieved data. First, for a generated response, we get the LLM Output shape using Vietoris Rips algorithm. Next, to make TAD scalable across a vast amount of logs, we start by partitioning data into $\sqrt{N}$ and do Wasserstein distance of the original output vs. the ablated batch output. We do this over all the transformer layers and aggregate the overall distance. For batches with largest gap and above, we proceed to Step-3 to analyze the remaining logs. Finally, we do one-by-one ablation with Wasserstein distance to trace the top attributed log(s).}
\label{fig:tad_process}
\end{figure*}

\section{Methodology}

In this section we instantiate the TDA pipeline discussed in Sec. \ref{sec: TDA_background} within the context of LLMs, and discuss the proposed methodology on using Wasserstein distance to track attribution of generations over the transformer layers.

\subsection{Mapping TDA concepts to LLMs}

\noindent \textbf{Token Embeddings as Metric Spaces}: In the LLM setting, each point $v_i$ in the point cloud corresponds to a \textit{vector representation (embedding)} of a token.  The space is $M = \mathbb{R}^n$, where $n$ is the model's hidden dimension, and the metric $d$ captures geometric dissimilarity between embeddings.  We adopt the $L_\infty$ distance, as can show salient directional differences in high-dimensional embedding spaces. 

\vspace{2pt}

\noindent \textbf{Simplicial Complexes to Encode Semantic Relationships:} We apply the Vietoris--Rips complex to the token embedding point cloud. At threshold $\varepsilon$, two token embeddings are connected by an edge whenever their $L_\infty$ distance is at most $\varepsilon$.  Higher-order simplices form when groups of embeddings are mutually within $\varepsilon$ of one another. Since no single $\varepsilon$ captures the full hierarchical organization of the embedding space, we construct the full Vietoris-Rips filtration over the token embeddings.  Each $K_{\varepsilon_i}$ represents the simplicial complex at scale $\varepsilon_i$, and the nested sequence tracks how semantic groupings form and merge as the threshold grows.

Filtrations can be driven by scalar functions defined on the embeddings. In our setting, a natural choice is the \textbf{sublevel-set filtration} parameterized by \textbf{Layer-wise Hidden States}. The filtration captures which tokens are ``attended to'' at each stage. This construction connects topological analysis directly to model-internal signals, making the filtration sensitive to saliency and anomalous patterns in LLM representations.

\noindent \textbf{Homology Groups for Topological Invariants of the Embedding Space:} At each scale $\varepsilon_i$, we compute the homology groups of the Vietoris-Rips complex built over the token embeddings: $H_0$ reflects how embeddings group into coherent \textbf{semantic regions (clusters)} at a given scale. $H_1$ captures \textbf{cycles} in the embedding space, which may correspond to gaps or circularly-related semantic concepts in representation. Higher-dimensional groups $H_k$ ($k \geq 2$) encode more abstract multi-way relational structures among embeddings. These homological invariants provide a compact summary of the global geometry of the embedding space at each scale.

\noindent \textbf{Persistent Homology to Track Semantic Structure Across Scales:} Applying persistent homology to the token embedding filtration allows for a complete record of which semantic structures appear and disappear as $\varepsilon$ grows.  A \textbf{birth} event in $H_0$ corresponds to a new isolated connected component emerging at scale $\varepsilon_b$. A \textbf{death} event in $H_0$ corresponds to two connected components merging into a broader group at scale $\varepsilon_d$. Births and deaths in $H_1$ correspond to the formation and filling of cycles of embeddings, which may indicate conceptual relationships. In this work, our primary focus is on $H_0$ homology, but we will provide $H_1$ results in Appendix \ref{appendix: h1_resutls}.
The \textbf{persistence} $\operatorname{pers}(\alpha) = \varepsilon_d - \varepsilon_b$ of a semantic feature $\alpha$ quantifies its stability. Features with high persistence represent robust and meaningful patterns in the embedding space, such as well-defined topics, while features with low persistence could arise from noise or incidental embedding proximity. The multiset of all such persistence values forms a \textbf{persistence diagram} $\operatorname{Dgm}(X) \subset \mathbb{R}^2$, where each feature $\alpha$ contributes a point $(\varepsilon_b, \varepsilon_d)$. The structural similarity between two such diagrams can then be quantified via the \textbf{Wasserstein distance} (refer to Equ. \ref{equ: wasserstein}). Therefore, one could ask: ``How can these topological descriptors be leveraged to systematically track the attribution and mapping of semantic domains across embedding spaces?'' We discuss this in the following section.

\subsection{Topological Attribution Distance (TAD)}

We now turn to the problem of attributing model generations to individual incident log segments. Our approach uses persistent homology to quantify how each retrieved log segment influences the topological structure of the model’s response. 
To quantify how different RAG segments influence model output, we must first ground our approach in the fundamental properties of decoder-only LLMs: autoregressive token generation. Since each generated token depends exclusively on all preceding tokens in the context window, a critical implication follows: every element in the prompt contributes to shaping the response geometry with varying degrees of causal influence. Therefore, the positional placement, sequential ordering, and compositional structure of RAG segments within the context window can produce measurable effects on model output.

To comprehensively measure how a given set of retrieved logs contributes to the final response, the methodology requires full context concatenation: all segments are appended sequentially into a single context along with the generated response. This design ensures that the causal relationship between each segment and the model's output can be traced without ambiguity. By doing so, segment-level contribution can be isolated through controlled ablation: systematically excluding individual segments across experimental runs while keeping all other variables constant. Together, these two mechanisms (i.e., full concatenation and targeted ablation) form the basis for attributing response content back to its originating context.

To get the topological mapping of a model's output with each log segment contributing to the output, we compute the $(\infty, 1)$-Wasserstein distance (refer to Equ. \ref{equ: wasserstein}) of H0 homology between their corresponding persistence diagrams on each layer of an LLM. The reason for H0 homology is that it is more computationally efficient (see Sec. \ref{sec: complexity}), and prior work has shown H0's effectiveness on geometric reorganization tasks \cite{yang2026persistent}. To capture an overall representation, we aggregate the Wasserstein values over the layers to obtain the attribution of each log segment. This is due to the nature of transformers that every layer correspond to the final decision, as also discussed by prior work \cite{gardinazzi2025persistent}. The following is the proposed TAD metric:

\begin{equation}
    \label{equ: tad_total}
  \mathcal{TAD}(P,\, Q)
  \;=\;
  \sum_{\ell=1}^{L}
  W_1^{(\infty)}\!\left(\operatorname{Dgm}_0(P_\ell),\, \operatorname{Dgm}_0(Q_\ell)\right)
\end{equation}

where $Dgm_0$ represents the H0 homology, and $P_l$ and $Q_l$ are persistent diagrams of the ablated context and the full context on each layer, respectively. This proposed extension can provide attribution beyond token-level impact, as it can provide explainability at segment-level. The key insight is that each log segment that most strongly influences the output, bears representational similarity to the output itself in the feature space. Therefore, ablating these influential segments accordingly should produce substantially larger Wasserstein distances, providing a direct measure of their contribution.

Figure \ref{fig:tad_process} demonstrates our proposed methodology. First, we characterize the topological structure of the generated output by constructing a Vietoris-Rips complex over its representation and extracting H0 persistent homology with the full context + the generated response. Next, since per-log ablation is computationally prohibitive at scale, we adopt a \emph{screen-then-confirm} attribution strategy. In the \emph{screen} stage, the input log sequence is partitioned into approximately $\sqrt{N}$ batches. Each batch is ablated in turn, and the resulting perturbation to output shape is quantified via the Wasserstein distance between the persistence diagrams of the original and ablated LLM response, computed independently at each transformer layer and aggregated across layers into a single distance measure per batch. The batch containing the largest aggregate distance is then selected for finer-grained analysis. In the \emph{confirm} stage, we perform single-log ablation restricted to this high-attribution batch, again computing layer-wise Wasserstein distances between resulting persistence diagrams to identify and rank the individual log entries most responsible for shaping the model's output geometry. In \emph{Algorithm 1}, we demonstrate the full detail process for the proposed TAD methodology.

  \begin{algorithm}[!tb]                                                                                           
  \caption{TAD: Topological Attribution Distance}                                                                
  \label{alg:tad}                                                                                                
  \begin{algorithmic}[1]                                                                                         
  \Require Context $C$ with logs $\mathcal{L}$, LLM $\mathcal{M}$ with $L$ layers,                               
           homology dimension $d$, min grouping size $\tau$ (default $9$)                                                                                
  \Ensure Set of spike logs $\mathcal{S} \subseteq \mathcal{L}$                                                  
  \State $R \gets \mathcal{M}(C)$ \Comment{Response, generated once and held fixed}                              
  \State $\mathcal{D}_{\text{base}}^{(1:L)} \gets \Call{Diagrams}{C, R, d}$                                      
  \If{$|\mathcal{L}| < \tau$} \Comment{Too small to screen; confirm every log}                                
      \State \Return \Call{Confirm}{$\mathcal{L}, C, R, \mathcal{D}_{\text{base}}^{(1:L)}$}                      
  \EndIf                                                                                                         
  \State $\mathcal{H} \gets \Call{Screen}{\mathcal{L}, C, R, \mathcal{D}_{\text{base}}^{(1:L)}}$                 
  \State $\mathcal{S} \gets \Call{Confirm}{\mathcal{H}, C, R, \mathcal{D}_{\text{base}}^{(1:L)}}$                
  \State \Return $\mathcal{S}$                                                                                
  \Statex\hrulefill                                                                                              
\Procedure{Diagrams}{$C', R, d$} \Comment{Only response tokens}
    \State $X^{(1:L)} \gets \Call{HiddenStates}{\mathcal{M}, C' \oplus R}$
    \State \Return $\big(\Call{VietorisRips}{X^{(\ell)}\!\restriction_R, d}\big)_{\ell=1}^{L}$
\EndProcedure                                                                     
  \Statex\hrulefill                                                                                              
  \Procedure{Screen}{$\mathcal{L}_{\text{pool}}, C, R, \mathcal{D}_{\text{base}}^{(1:L)}$}                       
      \State $k \gets \lceil \sqrt{|\mathcal{L}_{\text{pool}}|} \rceil$                                          
      \State $\mathcal{G} \gets \Call{Partition}{\mathcal{L}_{\text{pool}}, k}$                                  
      \For{each group $G_i \in \mathcal{G}$}                                                                     
          \State $C' \gets \Call{Ablate}{C, G_i}$ \Comment{Only the context changes}                             
          \State $\mathcal{D}^{\prime(1:L)} \gets \Call{Diagrams}{C', R, d}$                                     
          \State $\Delta_i \gets \Call{TAD}{\mathcal{D}^{\prime}, \mathcal{D}_{\text{base}}^{}}$            
      \EndFor                                                                                                    
      \State $\mathcal{L}_{\text{hot}} \gets \bigcup_{G \in \mathcal{G}_{\text{hot}}} G$                         
      \If{$|\mathcal{L}_{\text{hot}}| \geq \tau$} \Comment{re-partition}                    
          \State \Return \Call{Screen}{$\mathcal{L}_{\text{hot}}, C, R, \mathcal{D}_{\text{base}}^{(1:L)}$}                                                  
      \EndIf                                                                                                     
      \State \Return $\mathcal{L}_{\text{hot}}$                                                                  
  \EndProcedure                                                                                                  
  \Statex\hrulefill                                                                                              
  \Procedure{Confirm}{$\mathcal{H}, C, R, \mathcal{D}_{\text{base}}^{(1:L)}$}                                    
      \For{each log $l \in \mathcal{H}$}                                                                         
          \State $C' \gets \Call{Ablate}{C, \{l\}}$                                                              
          \State $\mathcal{D}^{\prime(1:L)} \gets \Call{Diagrams}{C', R, d}$                                     
          \State $\Delta_l \gets \Call{TAD}{\mathcal{D}^{\prime}, \mathcal{D}_{\text{base}}^{}}$            
      \EndFor                                                                                                    
      \State \Return \Call{FlagSpikes}{$\mathcal{H}, \Delta$}                                                    
  \EndProcedure                                                                                                  
  \Statex\hrulefill                                                                                              
  \Procedure{FlagSpikes}{$\mathcal{X}, \Delta$}                                                                  
      \State Sort $\Delta$ descending with order $\sigma$                                                        
      \State $\text{gaps}[i] \gets \Delta_{\sigma[i]} - \Delta_{\sigma[i+1]}$                                    
      \State $c \gets \arg\max_i \text{gaps}[i]$                                                                 
      \State \Return $\{\mathcal{X}_{\sigma[i]} : i \leq c \land \text{gaps}[c] > 0\}$                           
  \EndProcedure                                                                                                  
  \end{algorithmic}                                                                                              
  \end{algorithm}       

\section{Discussion on Complexity}
\label{sec: complexity}

The cost of TAD decomposes into two stages: transformer inference and persistent homology. Let $N$ denote the number of candidate log entries in the context, $T$ the number of response tokens in the point cloud, $d$ the hidden dimension of the model, and $L$ the number of transformer layers.

\noindent \textbf{Inference cost.} TAD is attribution-by-ablation, therefore, the complexity is the number of forward passes $M$ issued to the model. A purely linear case, in which each log is ablated individually, requires $M = N + 1$ passes (one baseline plus one per log), which scales linearly in context length and becomes prohibitive for long Agentic traces. Our proposed screen-then-confirm procedure instead partitions the pool into $\lceil \sqrt{N} \rceil$ groups and recurses only on the groups flagged as spikes, which reduces the expected cost to $M = O(\sqrt{N})$ passes.

\noindent \textbf{Topological cost.} For each forward pass and each layer $\ell$, TAD builds a point cloud $X_\ell \in \mathbb{R}^{T \times d}$ from the response-token hidden states and computes the persistence diagram of its Vietoris--Rips filtration. Forming the pairwise distance matrix costs $O(T^{2} d)$. When restricted to $H_0$, only the $1$-skeleton is required: the $O(T^{2})$ edges are sorted in $O(T^{2} \log T)$ time. The per-pass cost is therefore $O\!\left(L\, T^{2} (d + \log T)\right)$, i.e. quadratic in the number of response tokens and linear in network depth. Extending to $H_1$ is substantially more expensive, as it needs $2$-skeleton, which may contain $O(T^{3})$ triangles, and can have cubic complexity in the number of simplices. 
Overall, the end-to-end cost of TAD is thus $O\!\left(M \cdot L \cdot T^{2}(d + \log T)\right)$ in the $H_0$ homology, with $M = O(\sqrt{N})$ under adaptive screening, and it is the configuration we adopt throughout our experiments.

\section{Results}

In this section, we will introduce our dataset, experimental design, and TAD results for incident log analysis.

\subsection{Dataset Curation \& Attack Scenario}

We curated the dataset from logs collected during a real cyberattack executed in the Game of Active Directory (GOAD) environment \cite{goad} and monitored using the Wazuh security framework \cite{wazuh}. GOAD is an intentionally vulnerable Windows Active Directory environment designed for penetration testing and security training. The attack was conducted in October 2025 and during this timeframe, we identified 20 log entries as the ground-truth attack evidence. These entries correspond to events directly associated with the adversarial activities performed during the attack, while the remaining logs represent benign activity, or events not directly attributable to the attack.

Cybersecurity logs often contain long sequences of nearly identical events, such as brute-force attempts that differ only in timestamps, ports, or event identifiers. To reduce this redundancy, we applied Gestalt Pattern Matching to group consecutive logs with $\ge 90\%$ sequence similarity and represented each group with a single entry. We added a \texttt{how-many-consecutive} metadata field to record the number of original events represented by each compressed log. This preserves event frequency while preventing repetitive activity from dominating the analysis. After compression, the dataset contains 587 log entries. Therefore, the compression reduced the log volume while preserving the semantics and frequency of the log events. In Table \ref{tab:dataset_overview} and Table \ref{tab:dataset_statistics}, we demonstrate the details of our dataset. 

\begin{table}[ht]
\centering
\caption{Dataset overview. Each window is a slice of Windows event logs surrounding the attack logs. \#GT refers to the number of ground-truth attack-relevant logs.}
\label{tab:dataset_overview}
\small
\begin{tabular}{@{}rlllcc@{}}
\toprule
\textbf{W} & \textbf{Date} & \textbf{Start (UTC)} & \textbf{End (UTC)} &
\textbf{\#Logs} & \textbf{\#GT} \\
\midrule
1  & 2025-10-10 & 18:51:21 & 18:52:16 & 71 & 3 \\
2  & 2025-10-10 & 18:56:59 & 18:57:45 & 48 & 1 \\
3  & 2025-10-10 & 19:02:21 & 19:03:40 & 48 & 2 \\
4  & 2025-10-10 & 19:09:46 & 19:11:17 & 50 & 4 \\
5  & 2025-10-10 & 19:54:49 & 19:55:47 & 46 & 3 \\
6  & 2025-10-10 & 20:00:00 & 20:01:41 & 59 & 3 \\
7  & 2025-10-11 & 03:35:51 & 03:37:51 & 81 & 1 \\
8  & 2025-10-11 & 06:56:27 & 06:57:05 & 47 & 1 \\
9  & 2025-10-11 & 18:56:41 & 18:57:26 & 50 & 1 \\
10 & 2025-10-11 & 20:48:45 & 20:48:46 & 87 & 1 \\
\midrule
\multicolumn{4}{@{}l}{\textbf{Total} (10 windows)} & 587 & 20 \\
\bottomrule
\end{tabular}
\end{table}                                                                                                   
                                     
\begin{table}[ht]                                                                                               
\centering   
\caption{Ground-truth attack events categories of the dataset.}                                  
\label{tab:dataset_statistics}                                                                                            
\small                                                                                                         
\begin{tabular}{@{}lc@{}}                                                                                    
\toprule                                                                                                       
\textbf{Technique} & \textbf{Count} \\                                        
\midrule                                                                                                       
Pass-the-Hash (NTLM remote logon)    & 9 \\                                           
Registry modification (BAM, tool execution)  & 4\\                                          
Malicious service creation (PowerShell payload) & 2 \\                                      
PSEXESVC service installation         & 2 \\                                           
Credential brute force (logon failure)       & 1 \\                                           
Interactive logon, compromised credentials  & 1  \\                                           
Registry value integrity change       & 1 \\                                           
\midrule                                                                                                       
\textbf{Total} & \textbf{20} \\                                                             
\bottomrule                                                                                                    
\end{tabular}                                                                                                  
\end{table}  

\vspace{2pt}

\noindent \textbf{Attack Progression Summary:}
The attack consists of a multi-stage intrusion involving pass-the-hash authentication, lateral movement, remote execution, tool deployment, and defense evasion. On the GOAD environment \cite{goad}, the attacker operated from \texttt{NPC-PETYERBAELI} host and compromised the \texttt{robb.stark} account to gain access to several systems.

\begin{enumerate}
    \item \textbf{Credential Validation and Initial Access:}
    On October 10, at approximately 18:51, a logon attempt for the \texttt{robb.stark} account failed on \texttt{NPC-PETYERBAELI}. This was later followed by a successful elevated interactive logon on the same host. Within seconds, the attacker used the same account in a successful pass-the-hash authentication against \texttt{winterfell}. This sequence shows that the attacker had obtained the account's NTLM credential.

    \item \textbf{Lateral Movement and Service Deployment:}
    Between approximately 18:57-19:10, the attacker used pass-the-hash authentication to access \texttt{winterfell}, \texttt{kingslanding}, and \texttt{meereen}. Several authentication attempts used consecutive source ports, showing that the activity was carried out through an automated sequence. After successful NTLM authentication, the attacker deployed PowerShell-based services with AMSI-bypass functionality on \texttt{winterfell}. These services allowed remote command execution.

    \item \textbf{Remote Execution:}
    Between approximately 19:55 and 20:01, the attacker deployed the \texttt{PSEXESVC} service and continued pass-the-hash authentication from \texttt{NPC-PETYERBAELI}.

    \item \textbf{Post-Exploitation Tooling and Defense Evasion:}
    On October 11, the attacker deployed and executed additional post-exploitation tools across several systems. \texttt{COLIncrease.exe} was executed on \texttt{castelblack} under the \texttt{robb.stark} account. A separate copy of the executable was later executed again on \texttt{NPC-PETYERBAELI}, as shown by  a change in the BAM entry's checksum. The attacker also executed \texttt{PsExec64.exe} on \texttt{NPC-PETYERBAELI} to support further remote execution. Finally, \texttt{DefenderRemover.exe} was executed on \texttt{vdi-samwell-tar} to disable Windows Defender and reduce endpoint security protections.
\end{enumerate}

\noindent
Overall, the sequence shows that the attacker systematically expanded access across the environment and weakened security controls to support continued malicious activity.

\begin{table*}[t]
\centering
\caption{Performance comparison of TAD (H0 homology) and other attribution tracing methods across four LLMs. The Direct, Regular, and Indirect cases point to the difficulty-level of the response with respect to the ground-truth logs.}
\label{tab:tad_main_results}
\setlength{\tabcolsep}{5pt}
\renewcommand{\arraystretch}{1.02}
\scriptsize

\resizebox{\textwidth}{!}{
\begin{tabular}{
  ll
  cc
  !{\vrule width 0.5pt}
  cc
  !{\vrule width 0.5pt}
  cc
  !{\vrule width 0.5pt}
  cc
}
\toprule
\multirow{2}{*}{\textbf{Case}} &
\multirow{2}{*}{\textbf{Method}} &
\multicolumn{2}{c}{\textbf{Qwen3-4B}} &
\multicolumn{2}{c}{\textbf{Gemma-3-4B}} &
\multicolumn{2}{c}{\textbf{Qwen2.5-7B}} &
\multicolumn{2}{c}{\textbf{Granite-4.1-8B}} \\
\cmidrule(lr){3-4}
\cmidrule(lr){5-6}
\cmidrule(lr){7-8}
\cmidrule(lr){9-10}
& &
\textbf{Accuracy} & \textbf{F1-score} &
\textbf{Accuracy} & \textbf{F1-score} &
\textbf{Accuracy} & \textbf{F1-score} &
\textbf{Accuracy} & \textbf{F1-score} \\
\midrule

\multirow{6}{*}{\textbf{Direct}}
& LLM-as-judge      & 0.9693 & 0.6897 & 0.9046 & 0.2821 & 0.9796 & \bfseries{0.7500} & 0.9625 & 0.6333 \\
& In-line citation  & 0.9710 & 0.5143 & 0.9693 & 0.4000 & 0.9455 & 0.4074 & 0.9779 & \textbf{0.6829} \\
& Cosine similarity & 0.8058 & 0.2192 & 0.8058 & 0.2192 & 0.8058 & 0.2192 & 0.8058 & 0.2192 \\
& ROUGE-L           & 0.7990 & 0.1690 & 0.7990 & 0.1690 & 0.7990 & 0.1690 & 0.7990 & 0.1690 \\
& LEA               & 0.9199 & 0.4051 & 0.9199 & 0.4051 & 0.9199 & 0.4051 & 0.8535 & 0.2712 \\
\rowcolor{gray!10}
& \textbf{TAD (ours)}
                    & \textbf{0.9847} & \textbf{0.7097} 
                    & \textbf{0.9779} & \textbf{0.6061}
                    & \textbf{0.9830} & 0.6667
                    & \textbf{0.9830} & 0.6667  \\
\addlinespace[2pt]
\midrule

\multirow{6}{*}{\textbf{Regular}}
& LLM-as-judge      & 0.8790 & 0.3604 & 0.7274 & 0.1304 & 0.9267 & 0.4416 & 0.9659 & 0.6296 \\
& In-line citation  & 0.9727 & 0.6000 & 0.9063 & 0.2466 & 0.9489 & 0.3478 & 0.9659 & 0.6296 \\
& Cosine similarity & 0.9131 & 0.3544 & 0.9131 & 0.3544 & 0.9131 & 0.3544 & 0.9131 & 0.3544 \\
& ROUGE-L           & 0.6968 & 0.1010 & 0.6968 & 0.1010 & 0.6968 & 0.1010 & 0.6968 & 0.1010 \\
& LEA               & 0.8007 & 0.1583 & 0.8262 & 0.1774 & 0.8007 & 0.1583 & 0.7104 & 0.1414 \\
\rowcolor{gray!10}
& \textbf{TAD (ours)}
                    & \textbf{0.9847} & \textbf{0.7097} 
                    & \textbf{0.9727} & \textbf{0.5000}
                    & \textbf{0.9796} & \textbf{0.6000} 
                    & \textbf{0.9813} & \textbf{0.6452}  \\
\addlinespace[2pt]
\midrule

\multirow{6}{*}{\textbf{Indirect}}
& LLM-as-judge      & 0.7819 & 0.2099 & 0.7172 & 0.1170 & 0.8756 & 0.2626 & 0.6968 & 0.1275 \\
& In-line citation  & 0.9131 & 0.3377 & 0.8876 & \textbf{0.1951} & 0.9216 & 0.2333 & 0.8790 & 0.2198 \\
& Cosine similarity & 0.6048 & 0.1008 & 0.6048 & 0.1008 & 0.6048 & 0.1008 & 0.6048 & 0.1008 \\
& ROUGE-L           & 0.6542 & 0.1057 & 0.6542 & 0.1057 & 0.6542 & 0.1057 & 0.6542 & 0.1057 \\
& LEA               & 0.8041 & 0.0171 & 0.7836 & 0.0155 & 0.8041 & 0.0171 & 0.8041 & 0.0171 \\
\rowcolor{gray!10}
& \textbf{TAD (ours)}
                    & \textbf{0.9796} & \bfseries{0.6000} 
                    & \textbf{0.9438} & 0.1538 
                    & \textbf{0.9761} & \textbf{0.5625} 
                    & \textbf{0.9659} & \textbf{0.3750}  \\
\bottomrule
\end{tabular}

}
\end{table*}

\subsection{Experimental Design}

In real-world scenarios, where thousands or millions of log entries may be generated, providing all available logs directly to an LLM is often impractical due to context-window limitations, increased inference cost, and the presence of irrelevant information that can degrade response quality. A targeted investigation allows the model to focus on the subset of evidence most relevant to the analyst's hypothesis while reducing noise and computational overhead. Therefore, our experimental scenario is motivated by a targeted threat-investigation workflow. We assume that a security analyst has already developed a suspicion that a particular user account may have been compromised. The analyst subsequently queries the system using a prompt of the following form:

\begin{promptbox}

{\it \small Examine the logs recorded between [start time] and [end time] on [date]. Is there evidence of an attack in these logs? }

\end{promptbox}

\noindent This formulation bounds the investigation by date and time, reflecting a realistic post-alert workflow. Rather than performing unrestricted threat hunting, the model assesses whether the supplied logs support the analyst’s hypothesis.

Furthermore, directly comparing natural-language responses from multiple LLMs introduces confounding factors such as differences in wording, detail, structure, confidence, and event interpretation. To improve comparability, we assume that a canonical response derived from the ground-truth attack logs is already available and hold this response constant across all evaluated models. To curate these canonical responses, we used \textit{Claude-Opus-4.5} model to generate high-quality outputs that closely match with the ground-truth attack logs. This design constitutes a controlled experimental abstraction. In a real-world deployment, each model would generate its own investigation report, and the resulting conclusions could vary according to the model’s training data, reasoning capabilities, and contextual understanding. Therefore, the use of a canonical response improves experimental comparability, but does not evaluate the quality of the model-specific response generation. 

To evaluate TAD on different difficulty levels, we considered three scenarios: ``Direct'', ``Regular'', and ``Indirect''. ``Direct'' is for when the LLM significantly copies the words from the source logs. The ``Regular'' case is for when an LLM copies some keywords but gives a more general interpretation. The ``Indirect'' case is for when the LLM almost do not use any critical word that can map the response to the specific log(s) via a keyword-search. The reason for this is that we will show TAD is not based on finding exact keywords and putting higher attention to those. Instead, it is based on the overall interpretation of each log (as a segment).

Furthermore, we compare TAD against multiple baselines: LLM-as-a-judge, in-line citation, similarity-based measures (Cosine Similarity and ROUGE-L), and a token-level attribution technique (LEA). These baselines represent commonly used approaches to assess the provenance of LLM-generated responses. For the LLM-as-a-judge and in-line citation baselines, we use greedy-decoding to get deterministic and reproducible responses. Appendix \ref{appendix: llm_prompts} shows the engineered prompts for these methods. For token-level attribution, we adopt our previously proposed LEA metric \cite{fayyazi2025llm}, as it is based on linear dependency of input embeddings and is computationally efficient. Note that we do not consider other token-level attribution methods, such as SHAP, due to their high computational complexity. Finally, for the cosine embedding model, we used the \texttt{Alibaba-NLP/gte-modernbert-base} model, as it is well-suited for long-context semantic search (8,192 tokens). 

The LLMs used for analysis are: \texttt{Qwen3-4B, Gemma3-4B, Qwen2.5-7B, and Granite4.1-8B}. The choice of these models are due to their long context window (a minimum of 128K context window) to handle large security logs and the ability to follow-instructions well. It is worth noting that we conducted our experiments on a system equipped with two Intel Xeon E5-2650 CPUs, 256 GB of RAM, and a NVIDIA Tesla L40S GPU. The code \& data is available at: \url{https://github.com/RezzFayyazi/TAD}.

\subsection{TAD for Incident Log Analysis}
\label{sec: tad_incident_log}

As discussed in the previous section, we compare TAD with different baselines under three settings, \emph{direct}, \emph{regular}, and \emph{indirect}, which differ in how openly the response refers to the logs. All methods use the same selection rule: we sort the candidates by score, find the largest gap between consecutive scores, and return everything above it. We do this because an analyst does not know in advance how many logs influenced a response. A fixed top-$k$ cutoff would either cut off real contributors or add noise, whereas the gap rule lets the scores set their own threshold.

Table \ref{tab:tad_main_results} demonstrates the results. As can be seen, TAD obtains the best accuracy and F1-score in almost all three settings, although the margin differs. On average, for the ``Direct'', ``Regular'', and ``Indirect'' cases, TAD outperforms the closest baseline by \textit{1.60\%}, \textit{3.11\%}, and \textit{6.63\%} in accuracy, and by \textit{7.35\%} \textit{15.77\%}, and \textit{17.63\%} margin in F1-score, respectively. Note that the margin ranges from \textit{1.60\%} to \textit{6.63\%} in accuracy, and from \textit{7.35\%} to \textit{17.63\%} in F1-score, with the ``Indirect'' case exhibiting the widest margin. This is because other baselines perform better when there are high-token overlaps from the logs and the response, and perform poorly when the response contains little-to-none shared tokens with the logs.
It is worth noting that in Appendix \ref{appendix: h1_resutls}, we provide TAD results based on H1 homology, and in Appendix \ref{appendix: examples}, we provide a qualitative example along with some specific examples on how TAD traces a log that contributed the most to the LLM response.

Moreover, based on our observation, the baselines indicated a clear pattern: Every baseline reaches high recall but much lower precision: it finds the influential logs, but only by returning a large set in which they sit among many false positives. TAD does this differently by reaching high Precision for the ``Direct'', ``Regular'', and ``Indirect'' cases with 94.23\%, 86.90\%, and 57.70\% respectively. On average, for the ``Direct'', ``Regular'', and ``Indirect'' cases, TAD outperforms the closest baseline by \textit{39.47\%}, \textit{47.46\%}, and \textit{40.88\%} margin in Precision, respectively. This matters in practice, as in the real-world, an analyst has limited time and a pool of thousands of logs. Therefore, a method that returns so many candidates (i.e., high recall) to reveal a few real ones, restores the manual work that automatic attribution is meant to remove. 
In addition, by looking at the results, one might ask: ``Why does the quality of the F1 score differ for each LLM?''. The reason is the \emph{learned representation of each LLM individually}. The learned representation matters on how each LLM connect tokens as a segment in the embedding space, as discussed by Rottach et al. \cite{rottach2025topology}. This means that if an LLM learned about cybersecurity more than the other, it naturally connects tokens in a more complex way. Therefore, an LLM like \textit{Gemma3-4B} performing worse than other models in our experiments, demonstrate that this model has not properly learned a mapping between security logs and the response.

Furthermore, the selected baselines perform worse for different reasons. First, LLM-based self-assessment is sensitive to prompt design, meaning that different engineered prompts can give different attribution judgments. This variability arises because the decision comes from the model's own reasoning over a follow-up prompt, rather than TAD's mathematically traceable internal property of the original context (that caused the LLM generation). Second, similarity and token-level methods fail because incident logs share highly overlapping tokens, causing the log that shaped the response almost identical to its near-duplicates under cosine similarity. TAD avoids both problems by measuring what the model does with the context (for a generated response) rather than how the context is similar to the response. Removing an influential log causes a large, clearly separated spike in the distance between layer-wise persistence diagrams. Because the response is fixed and the computation is deterministic, the result is reproducible, traceable, and reflects how the model maps context to response. Future work could investigate the importance of individual layers on the generation and explore scalable variants of the Vietoris-Rips construction. We see this as a step toward transparent and verifiable attribution that allows analysts to inspect and validate the evidence, rather than simply trust the model’s output.

\section{Conclusion}

We introduced Topological Attribution Distance (TAD) to capture the global geometric shape of an LLM output and its changes against its retrieved context. In other words, we show that when the embedding of a retrieved context (e.g., security logs) drastically changes the geometry of the model's response, this serves as robust verification of evidence, and can be traced. Therefore, we designed TAD powered by segment-level ablation attribution to investigate incident logs of an actual cyberattack, and we demonstrated how TAD finds the most attributed logs to the LLM output adaptively. TAD can provide an explainable and trustworthy tracing based on each LLM's hidden state to understand how different RAG segments geometrically influence the model generation, and provide evidence verification in cybersecurity operations and Agentic-AI workflows.

\section*{Acknowledgments}

This material is based upon work supported by the National Science Foundation under Grant No. 2344237 and No. 2502341. The authors gratefully acknowledge Dr. Justin Pelletier and Mr. Forrest Fuqua for their valuable contribution to the curation of the dataset.

\section*{AI Disclosure}
We used ChatGPT to assist with sentence-level editing and grammar checking. The tool also contributed to the aesthetic refinement of Fig. 1 and Fig. 3, and prompt-boxes. The authors independently reviewed and verified the accuracy, originality, and integrity of all content, including the cited references.

\bibliography{REFERENCES}
\bibliographystyle{IEEEtran}

\appendix

\subsection{TAD results based on H1 homology}
\label{appendix: h1_resutls}

In Table \ref{tab:tad_h1_results}, we show the TAD values based on H1 homology. As can be seen, the results are generally competitive in both settings, with H1 being higher in F1-score for the ``Indirect'' case for all four models. In addition, H1 results are also slightly higher for larger models, which could indicate that H1 can capture more abstract or higher-order representations as model capacity increases. However, due to H1 being computationally more expensive (cubic) than H0 (quadratic), we used H0 as the default setting for TAD. 

\begin{table*}[!ht]
\centering
\caption{TAD comparison based on H0 \& H1 homology.}
\label{tab:tad_h1_results}
\setlength{\tabcolsep}{4pt}
\renewcommand{\arraystretch}{1.2}
\scriptsize

\resizebox{\textwidth}{!}{%
\begin{tabular}{
  ll
  cc
  !{\vrule width 0.5pt}
  cc
  !{\vrule width 0.5pt}
  cc
  !{\vrule width 0.5pt}
  cc
}
\toprule
\multirow{2}{*}{\textbf{Case}} &
\multirow{2}{*}{\textbf{Method}} &
\multicolumn{2}{c}{\textbf{Qwen3-4B}} &
\multicolumn{2}{c}{\textbf{Gemma-3-4B}} &
\multicolumn{2}{c}{\textbf{Qwen2.5-7B}} &
\multicolumn{2}{c}{\textbf{Granite-4.1-8B}} \\
\cmidrule(lr){3-4}
\cmidrule(lr){5-6}
\cmidrule(lr){7-8}
\cmidrule(lr){9-10}
& &
{\textbf{Accuracy}} & {\textbf{F1-score}} &
{\textbf{Accuracy}} & {\textbf{F1-score}} &
{\textbf{Accuracy}} & {\textbf{F1-score}} &
{\textbf{Accuracy}} & {\textbf{F1-score}} \\
\midrule

\multirow{1}{*}{\textbf{Direct}}
& \textbf{TAD (H0)}
                    &  \textbf{0.9847} & 0.7097 
                    &  \textbf{0.9779} & \textbf{0.6061}
                    &  \textbf{0.9830} & 0.6667 
                    &  \textbf{0.9830} & 0.6667  \\

\addlinespace[2pt]

& \textbf{TAD (H1)}
                    & \textbf{0.9847} &  \textbf{0.7273}
                    & 0.9710 &  0.5405
                    & \textbf{0.9830} &  \textbf{0.7059}
                    & \textbf{0.9830} &  \textbf{0.7222} \\
\addlinespace[2pt]
\midrule

\multirow{1}{*}{\textbf{Regular}}
& \textbf{TAD (H0)}
                    &  \textbf{0.9847} &  \textbf{0.7097} 
                    &  \textbf{0.9727} &  \textbf{0.5000}
                    &  0.9796 &  0.6000 
                    &  0.9813 &  0.6452 \\

\addlinespace[2pt]

& \textbf{TAD (H1)}
                    &  0.9830 &  0.6667 
                    &  0.9438 &  0.2667 
                    &  \textbf{0.9830} &  \textbf{0.6667} 
                    &  \textbf{0.9847} &  \textbf{0.7097} \\

\addlinespace[2pt]
\midrule

\multirow{1}{*}{\textbf{Indirect}}
& \textbf{TAD (H0)}
                    &  0.9796 &  0.6000
                    &  \textbf{0.9438} &  0.1538 
                    &  \textbf{0.9761} &  0.5625 
                    &  \textbf{0.9659} &  0.3750  \\

\addlinespace[2pt]
                    
& \textbf{TAD (H1)}
                    &  \textbf{0.9813} &  \textbf{0.6452}
                    &  \textbf{0.9438} &  \textbf{0.2326}
                    &  \textbf{0.9761} &  \textbf{0.5882}
                    &  0.9608 &  \textbf{0.4103} \\
\bottomrule
\end{tabular}%
}
\end{table*}

\subsection{TAD Specific Examples}
\label{appendix: examples}
To demonstrate in more detail on how TAD actually attributes the logs to the response, we provided a detailed qualitative example in Figure \ref{fig:tad-qualitative-example}. As can be seen, TAD correctly attributes log-35 with the highest contribution into the LLM's response, with a total Wasserstein of 1951.13 and containing the largest gap (1162.72) compared to the other logs.

In addition, we provided three specific examples for the ``Direct'', ``Regular'', and ``Indirect'' cases. The results show how TAD correctly attributes all cases, and interestingly, for the indirect case, where there is no keyword similarity, TAD still managed to attribute to the correct log.

\begin{figure*}[!ht]
\centering

\begin{tcolorbox}[
    enhanced,
    width=\textwidth,
    colback=white,
    colframe=RuleGray,
    boxrule=0.4pt,
    arc=1.2mm,
    left=2mm,
    right=2mm,
    top=1.1mm,
    bottom=1.1mm,
    title=\textbf{TAD's Qualitative Example},
    fonttitle=\normalsize\bfseries,
    colbacktitle=HeaderBlack,
    coltitle=white,
    titlerule=0pt,
    toptitle=1mm,
    bottomtitle=1mm
]

\footnotesize
\setlength{\tabcolsep}{2.6pt}
\renewcommand{\arraystretch}{1.10}

\noindent\textbf{\underline{1. Analyst query}}

\vspace{0.5mm}

\begin{tcolorbox}[
    enhanced,
    width=\linewidth,
    colback=SoftGray,
    colframe=RuleGray,
    boxrule=0.3pt,
    arc=0.8mm,
    left=1.4mm,
    right=1.4mm,
    top=0.55mm,
    bottom=0.55mm
]
\footnotesize
Examine logs around 06:56--06:57 on 2025-10-11. Is there evidence of an attack in these logs?
\end{tcolorbox}

\vspace{1.2mm}

\noindent
\textbf{\underline{2. Candidate incident logs after the Screen process (refer to Algorithm~1)}}

\vspace{0.6mm}

\begin{tabularx}{\linewidth}{
    >{\centering\arraybackslash}p{0.05\linewidth}
    >{\raggedright\arraybackslash}X
    >{\centering\arraybackslash}p{0.12\linewidth}
    >{\centering\arraybackslash}p{0.14\linewidth}
    >{\centering\arraybackslash}p{0.09\linewidth}
}
\toprule

\rowcolor{SoftGray}
\textbf{Log} &
\textbf{Event} &
\textbf{State} &
\textbf{Total-Wasserstein} &
\textbf{Gap} \\

\midrule

31 &
BAM registry checksum changed for \texttt{powershell.exe}. &
\textcolor{MidGray}{Benign} &
\textcolor{MidGray}{\textbf{682.54}} &
\textcolor{MidGray}{\textbf{0.0}} \\

\rowcolor{SoftGray}
32 &
BAM registry checksum changed for \texttt{cmd.exe}. &
\textcolor{MidGray}{Benign} &
\textcolor{MidGray}{\textbf{708.30}} &
\textcolor{MidGray}{\textbf{25.75}} \\

29 &
BAM registry checksum changed for \texttt{ApplicationFrameHost.exe}. &
\textcolor{MidGray}{Benign} &
\textcolor{MidGray}{\textbf{710.32}} &
\textcolor{MidGray}{\textbf{2.03}} \\

\rowcolor{SoftGray}
34 &
BAM registry entry added for \texttt{rundll32.exe}. &
\textcolor{MidGray}{Benign} &
\textcolor{MidGray}{\textbf{749.56}} &
\textcolor{MidGray}{\textbf{39.24}} \\

30 &
BAM registry checksum changed for \texttt{windows.immersivecontrolpanel}. &
\textcolor{MidGray}{Benign} &
\textcolor{MidGray}{\textbf{782.09}} &
\textcolor{MidGray}{\textbf{32.53}} \\

\rowcolor{SoftGray}
33 & 
BAM registry entry added for temporary \texttt{Un.exe}. &
\textcolor{MidGray}{Benign} &
\textcolor{MidGray}{\textbf{788.61}} &
\textcolor{MidGray}{\textbf{6.52}} \\

\rowcolor{TADgreenRow}
\textbf{\textcolor{TADgreen}{35}} &
\textbf{BAM registry entry added for \texttt{PsExec64.exe} in \texttt{robb.stark}'s Downloads.} &
\textbf{\textcolor{TADgreen}{Malicious}} &
\textbf{\textcolor{TADgreen}{\textbf{1951.33}}} &
\textbf{\textcolor{TADgreen}{\textbf{1162.72}}} \\

\bottomrule
\end{tabularx}

\vspace{1.5mm}

\noindent
\textbf{\underline{3. LLM response (Granite-4.1-8B) for the regular case}}

\vspace{0.5mm}

\begin{tcolorbox}[
    enhanced,
    width=\linewidth,
    colback=SoftGray,
    colframe=RuleGray,
    boxrule=0.3pt,
    arc=0.8mm,
    left=1.4mm,
    right=1.4mm,
    top=0.7mm,
    bottom=0.7mm
]
\scriptsize
``NPC-PETYERBAELI records a Background Activity Moderator entry at 06:57 showing \texttt{PsExec64.exe} ran from \texttt{robb.stark}'s Downloads folder. PsExec spawns processes on remote machines, and a copy staged in a download folder rather than an approved install path indicates an operator preparing to execute code on other hosts.''
\end{tcolorbox}

\vspace{1.2mm}

\noindent
\textbf{\underline{4. Evidence localization}}

\vspace{0.6mm}

\begin{tabularx}{\linewidth}{
    >{\raggedright\arraybackslash}p{0.19\linewidth}
    >{\raggedright\arraybackslash}p{0.13\linewidth}
    >{\centering\arraybackslash}p{0.13\linewidth}
    >{\raggedright\arraybackslash}X
}
\toprule

\rowcolor{SoftGray}
\textbf{Family} &
\textbf{Method} &
\textbf{Result} &
\textbf{Operational consequence} \\

\midrule

\rowcolor{TADgreenRow}
\textbf{Segment-level attribution} &
\textbf{TAD (ours)} &
\cmark\ \textbf{Log 35} &
\textbf{Uniquely identifies the log mostly influencing the response geometry} \\

\bottomrule
\end{tabularx}

\end{tcolorbox}

\caption{
TAD qualitative example to demonstrate how Wasserstein distance measures each log's effect on the response geometry.
}
\label{fig:tad-qualitative-example}

\end{figure*}

\begin{mybox}{Direct Case -- Gemma3-4B -- TAD}

\textbf{\underline{LLM Output}}

\begin{tcolorbox}[
    colback=gray!6,
    colframe=gray!35,
    boxrule=0.2pt,
    arc=1.5mm,
    left=3mm,
    right=3mm,
    top=2mm,
    bottom=2mm
]

{\it \small At \texttt{03:36:16.781}, agent \texttt{castelblack} (\texttt{192.168.56.22}) logged ``Registry Value Entry Added to the System.'' A registry value associated with \texttt{COLIncrease.exe}, located on user \texttt{robb.stark}'s Desktop, was added under the Windows \texttt{bam\textbackslash State\textbackslash UserSettings} registry path in \texttt{scheduled} mode.}

\end{tcolorbox}

\tcblower

\textbf{\underline{TAD's Traced Log}}

\begin{lstlisting}[style=logstyle]
log_id=iLZX0ZkBDpTsYhzSRWvf
timestamp=2025-10-11T03:36:16.781Z
agent_name=castelblack
agent_ip=192.168.56.22
rule_description="Registry Value Entry Added to the System"
target_user=null
target_domain=null
full_log="Registry Value '[x32] HKEY_LOCAL_MACHINE\System\CurrentControlSet\Services\bam\State\UserSettings\S-1-5-21-3733740424-\\
1936739333-3027215852-1115\\Device\HarddiskVolume1\Users\robb.stark\Desktop\COLIncrease.exe' added Mode: scheduled"
\end{lstlisting}

\end{mybox}

\begin{mybox}{Indirect Case -- Qwen3-4B -- TAD}

\textbf{\underline{LLM Output}}

\begin{tcolorbox}[
    colback=gray!6,
    colframe=gray!35,
    boxrule=0.2pt,
    arc=1.5mm,
    left=3mm,
    right=3mm,
    top=2mm,
    bottom=2mm
]

{\it \small
An execution ledger on this box shows that a well-known utility for launching programs on other boxes was run from a fetch folder in a staff profile. Storing
that kit in a personal folder rather than an approved software area, then running it, points to an outsider preparing to reach further boxes.
}

\end{tcolorbox}

\tcblower

\textbf{\underline{TAD's Traced Log}}

\begin{lstlisting}[style=logstyle]
log_id=B7YP0pkBDpTsYhzSDo05
timestamp=2025-10-11T06:57:04.447Z
agent_name=npc-petyerbaeli
agent_ip=192.168.56.154
rule_description="Registry Value Entry Added to the System"
target_user=null
target_domain=null
full_log="Registry Value '[x32] HKEY_LOCAL_MACHINE\System\CurrentControlSet\Services\bam\State\UserSettings\S-1-5-21-3733740424-1936739333-3027215852
-1115\\Device\HarddiskVolume1\Users\robb.stark\Downloads\PsExec64.exe' ..."
\end{lstlisting}

\end{mybox}

\begin{mybox}{Regular Case -- Granite4.1-8B -- TAD}

\textbf{\underline{LLM Output}}

\begin{tcolorbox}[
    colback=gray!6,
    colframe=gray!35,
    boxrule=0.2pt,
    arc=1.5mm,
    left=3mm,
    right=3mm,
    top=2mm,
    bottom=2mm
]

{\it \small
WINTERFELL accepts a successful remote logon for \texttt{robb.stark} at 18:57, authenticated with NTLM from \texttt{NPC-PETYERBAELI}. Presenting a hash rather
than a typed password, and doing so against a domain controller from an ordinary workstation, is the signature of a replayed credential. This is unauthorized lateral movement rather than legitimate remote administration.
}

\end{tcolorbox}

\tcblower

\textbf{\underline{TAD's Traced Log}}

\begin{lstlisting}[style=logstyle]
log_id=FbV8z5kBDpTsYhzSINoU
timestamp=2025-10-10T18:57:20.490Z
agent_name=winterfell
agent_ip=192.168.56.11
rule_description="Successful Remote Logon Detected - User:\robb.stark - NTLM authentication, possible pass-the-hash attack... Verify that NPC-PETYERBAELI is allowed to perform RDP connections"
target_user=robb.stark
target_domain=NORTH
system_message="\"An account was successfully logged on. ... Security ID: S-1-5-21-37337...1115 Account Name: robb.stark Account Domain: NORTH ... Workstation Name: NPC-PETYERBAELI Source Network Address: 192.168.56.154 ... Package Name (NTLM only): NTLM V2 Key Length....\""
\end{lstlisting}

\end{mybox}

\subsection{LLM Self-Eval Baseline Prompts}
\label{appendix: llm_prompts}

Below we provide the prompts used for the LLM-based self-assessment baselines (LLM-as-Judge and in-line Citation). These prompts are designed to evaluate the extent to which the LLM can map the generated responses to the source logs and attribute their outputs to the specific retrieved segments that support them. We used greedy decoding for these cases to get deterministic responses. However, a key limitation of LLM-based judges is their sensitivity to engineering of prompts, as variations in prompt wording or structure can lead to different evaluation outcomes. TAD, on the other hand, does not rely on a secondary set of evaluation prompts to trace the logs and will only look at the response geometry.

\begin{mybox}{In-line Citation Prompt}
\begin{tcolorbox}[
    colback=gray!6,
    colframe=gray!35,
    boxrule=0.2pt,
    arc=1.5mm,
    left=3mm,
    right=3mm,
    top=2mm,
    bottom=2mm
]

{\itshape\small

\{formatted\_logs\}

\vspace{4pt}

Here is your response based on these logs:

\vspace{4pt}

\texttt{"""}\{response\}\texttt{"""}

\vspace{4pt}

For every sentence in the response, add inline citations to ALL logs that support that sentence.

\vspace{4pt}

- If multiple logs support a sentence, cite all of them.

\vspace{4pt}

- Do not leave factual claims uncited.

\vspace{4pt}

Output format: Add citations with
\texttt{\textless\textless logN\textgreater\textgreater}
at the end of each sentence.

}
\end{tcolorbox}
\end{mybox}

\begin{mybox}{LLM-as-Judge Prompt}
\begin{tcolorbox}[
    colback=gray!6,
    colframe=gray!35,
    boxrule=0.2pt,
    arc=1.5mm,
    left=3mm,
    right=3mm,
    top=2mm,
    bottom=2mm
]

{\itshape\small

You are an expert at analyzing cybersecurity incident responses.

Below are security logs, each tagged with 
\texttt{\textless\textless logN\textgreater\textgreater}%
...\texttt{\textless\textless logN/\textgreater\textgreater}:

\vspace{4pt}

\{formatted\_logs\}

\vspace{4pt}

You already examined these logs and produced this response:

\vspace{4pt}

\texttt{"""}

\{response\}

\texttt{"""}

\vspace{4pt}

TASK: Identify which specific logs you relied upon to produce this response.

\vspace{4pt}

RULES:

\vspace{4pt}

1. Only list logs that directly contributed to the response's conclusions

\vspace{4pt}

2. Output ONLY the log tags that were used (e.g.,
\texttt{\textless\textless log5\textgreater\textgreater},
\texttt{\textless\textless log12\textgreater\textgreater})

\vspace{4pt}

3. Do not include logs that were present but not referenced in the response

\vspace{4pt}

Available log tags: \{available\_tags\}

\vspace{4pt}

OUTPUT FORMAT:

List the relevant log tags, one per line:

\vspace{4pt}

\texttt{\textless\textless logX\textgreater\textgreater}

\texttt{\textless\textless logY\textgreater\textgreater}

\vspace{4pt}

RELEVANT LOGS:

}
\end{tcolorbox}
\end{mybox}

\end{document}